\documentclass[final,5p,times,twocolumn,numbers]{elsarticle}

\usepackage[numbers]{natbib}
\usepackage[superscript]{cite}
\usepackage[hyphens]{url}
\usepackage{graphicx}
\usepackage{booktabs}
\usepackage{makecell}
\usepackage{afterpage}
\usepackage{caption}

\journal{...}

\begin{document}

\begin{frontmatter}




\title{In the Shadow of Smith’s Invisible Hand: Risks to Economic Stability and Social Wellbeing in the Age of Intelligence}

\author{Jo-An Occhipinti\textsuperscript{a,b,c,*}}
\cortext[cor1]{Corresponding author. Email: jo-an.occhipinti@sydney.edu.au}
\affiliation[first]{Brain and Mind Centre, University of Sydney, Camperdown, Australia}
\affiliation{Mental Wealth Initiative, University of Sydney, Camperdown, Australia}
\affiliation{Computer Simulation & Advanced Research Technologies (CSART), Sydney, Australia}

\author{William Hynes\textsuperscript{b,d,e}}
\affiliation{World Bank, Paris, France}
\affiliation{Santa Fe Institute, Santa Fe, New Mexico, USA}

\author{Ante Prodan\textsuperscript{a,b,c,f}}
\affiliation{School of Computer, Data and Mathematical Sciences, Western Sydney University, Sydney, Australia}

\author{Harris A. Eyre\textsuperscript{b,g,h,i}}
\affiliation{Brain Capital Alliance, San Francisco, California, USA}
\affiliation{Baker Institute for Public Policy, Rice University, Houston, Texas, USA}
\affiliation{Meadows Mental Health Policy Institute, Dallas, Texas, USA}

\author{Roy Green\textsuperscript{j}}
\affiliation{University of Technology Sydney, Broadway, Sydney, Australia}

\author{Sharan Burrow\textsuperscript{k}}
\affiliation{Visiting Professor in Practice, London School of Economics Grantham Institute, London, UK}

\author{Marcel Tanner\textsuperscript{l,m}}
\affiliation{Swiss Academies of Arts and Sciences, Bern, Switzerland }
\affiliation{Swiss Tropical and Public Health Institute & University of Basel, Switzerland}

\author{John Buchanan\textsuperscript{n}}
\affiliation{Business School, University of Sydney, Sydney, Australia}

\author{Goran Ujdur\textsuperscript{a,b,c}}

\author{Frederic Destrebecq\textsuperscript{o}}
\affiliation{European Brain Council, Brussels, Belgium}

\author{Christine Song\textsuperscript{a,b}}

\author{Steven Carnevale\textsuperscript{p,q}}
\affiliation{Point Cypress Ventures, San Francisco, California, USA}
\affiliation{Mental Health Services Oversight & Accountability Commission, Sacramento, California, USA}

\author{Ian B. Hickie\textsuperscript{a,b}}

\author{Mark Heffernan\textsuperscript{r}}
\affiliation{Dynamic Operations, Sydney, Australia}





\begin{abstract}
Work is fundamental to societal prosperity and mental health, providing financial security, identity, purpose, and social integration. The emergence of generative artificial intelligence (AI) has catalysed debate on job displacement. Some argue that many new jobs and industries will emerge to offset the displacement, while others foresee a widespread decoupling of economic productivity from human input threatening jobs on an unprecedented scale. This study explores the conditions under which both may be true and examines the potential for a self-reinforcing cycle of recessionary pressures that would necessitate sustained government intervention to maintain job security and economic stability. A system dynamics model was developed to undertake ex ante analysis of the effect of AI-capital deepening on labour underutilisation and demand in the economy. Results indicate that even a moderate increase in the AI-capital-to-labour ratio could increase labour underutilisation to double its current level, decrease per capita disposable income by 26\% (95\% interval, 20.6\% - 31.8\%), and decrease the consumption index by 21\% (95\% interval, 13.6\% - 28.3\%) by mid-2050. To prevent a reduction in per capita disposable income due to the estimated increase in underutilization, at least a 10.8-fold increase in the new job creation rate would be necessary. Results demonstrate the feasibility of an AI-capital-to-labour ratio threshold beyond which even high rates of new job creation cannot prevent declines in consumption. The precise threshold will vary across economies, emphasizing the urgent need for empirical research tailored to specific contexts. This study underscores the need for governments, civic organisations, and business to work together to ensure a smooth transition to an AI-dominated economy to safeguard the Mental Wealth of nations.

\end{abstract}


\end{frontmatter}




\label{Introduction }
\section*{Introduction }
Quality work has long been recognized as a fundamental pillar of prosperity and mental health, providing not just financial security and a sense of identity and purpose, but fostering social connection and integration – all of which are fundamental to a thriving society \cite{RN1, RN2, RN3}. However, the fabric of work has been undergoing significant change over the past decades with a distinct shift towards more precarious employment along with work intensification \cite{RN4,RN5}, raising concerns about the impact of this shift on mental health and societal wellbeing \cite{RN6,RN7}. Particularly affected by these changes are young people. The International Labour Organization reported in 2022 that those aged 15-24 years faced significant challenges in securing and retaining quality employment, experienced an unemployment rate three times that of adults, and more than one in five were not engaged in education, employment or training \cite{RN8}. The shift towards greater labour market flexibility and job insecurity over the last four decades has been largely tolerated and justified as necessary for continued economic growth and prosperity. However, the adverse impacts of this shift on mental health, especially among young people, is a significant concern. Financial hardship, job quality and job insecurity are well-documented risk factors for mental health issues \cite{RN9,RN10, RN11, RN12}. Against a backdrop of declining productivity growth, and evidence that reduced job security and increasing demands for worker productivity paradoxically undermines productivity \cite{RN6}, efforts to safeguard worker wellbeing have moved into focus. This focus has served as a counterweight against further deterioration in job security and quality. However, the emergence of generative artificial intelligence (AI) poses a substantial threat to this balancing force.    

\begin{figure}[!htb]
\centering
\includegraphics[width=0.48\textwidth]{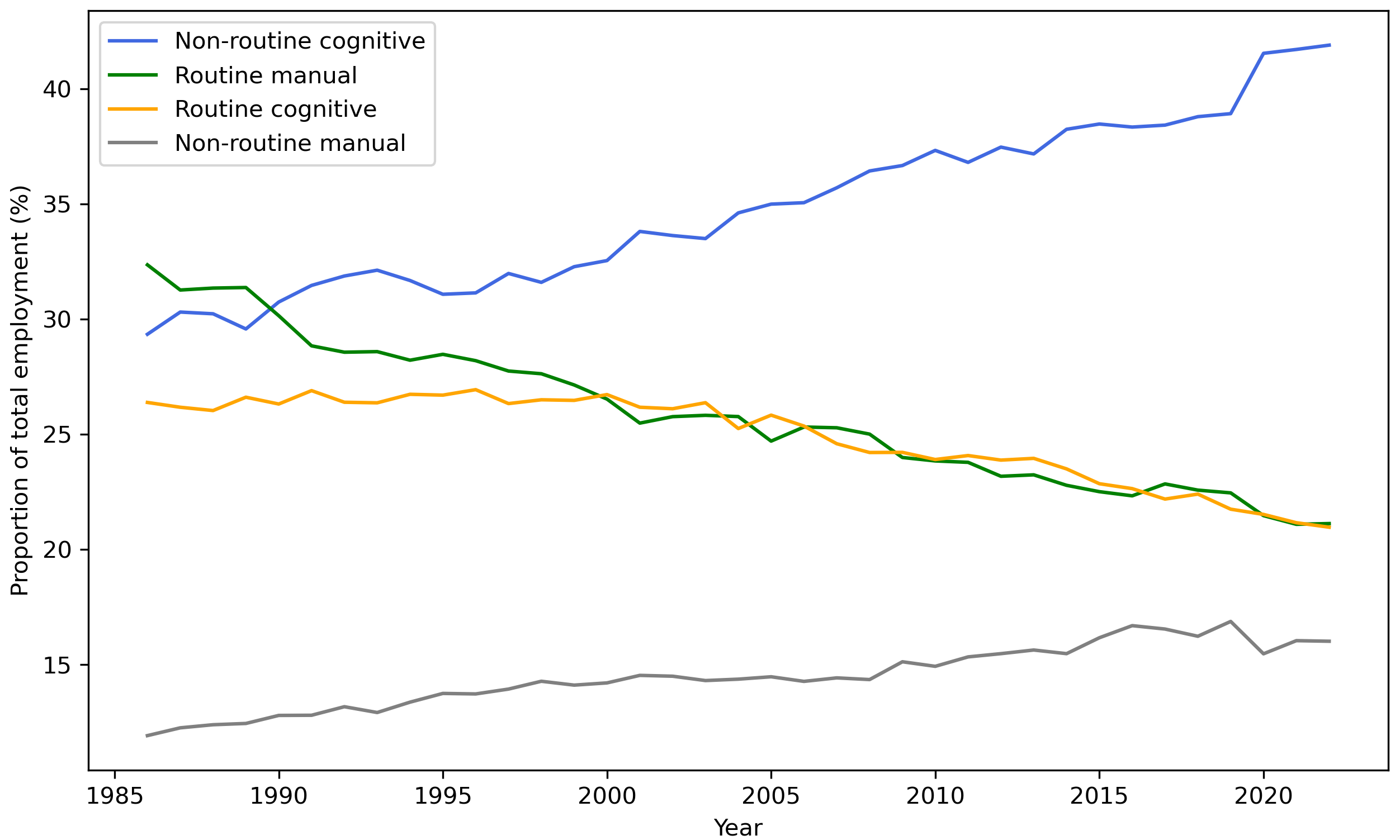}
\caption{Share of employment by type of occupation, Australia, 1986 – 2022 (August) (reprinted with permission. Source: Borland, J. \& Coelli, M. The Australian labour market and IT-enabled technological change. Working Paper No. 01/23. Melbourne Institute of Applied Economic \& Social Research, Melbourne, Australia, 2023).}
\label{fig:image1}
\end{figure}

Generative AI is not simply automation, it is an intelligence; it stands apart from previous technological advancements in its ability to undertake routine and non-routine cognitive tasks across a broad range of disciplines and is on a trajectory to achieving beyond human capabilities. This capability, while currently in its infancy, has the potential to substantially deepen the decoupling of economic productivity from labour input \cite{RN13,RN14}, putting at risk a large proportion of jobs once considered safe from automation, and profoundly reshaping labour market dynamics \cite{RN15, RN16}. Past technological advancements have led to the automation of routine manual and cognitive tasks, which has decreased the demand for work requiring these skills. Concurrently, there has been an expansion of opportunities for non-routine cognitive tasks (Figure \ref{fig:image1}), typically associated with higher wages. This shift has contributed to the growth of the middle class and has helped sustain economic demand. However, it is precisely this substantial segment of non-routine cognitive or professional jobs that is now facing potential disruption from the capabilities of generative AI. While most agree on the likelihood of short-term disruption as generative AI is adopted across the economy, opinions diverge on its overall impact on job displacement. Some argue that generative AI will create new jobs and industries (requiring human labour) that will offset displaced jobs consistent with the experiences with previous technological developments, and bring with it advances in healthcare, cheaper goods and services, more efficient allocation of resources, educational and reskilling opportunities, innovation, and new investments, products and services \cite{RN17}.  Others warn that the scale and nature of job displacement will be unprecedented, that generative AI is cheaper and more productive than human capital, it can be ‘upskilled’ far more efficiently than humans, and that the resulting labour underutilisation, downward pressure on wages, and middle class contraction will bring forth social and economic instability \cite{RN15}. We explore through dynamic modelling the conditions under which both may be true. 

The capital-to-labour ratio is an economic indicator that measures the amount of capital (such as physical infrastructure, machinery, or technology) that is available per worker in the production process. It serves as a key determinant of productivity, providing insights into labour efficiency and the capital intensity of an economy or industry \cite{RN18}. However, in the Age of Intelligence, while significant investments in AI-capital (increasing the numerator) is similarly expected to drive up productivity, it is also likely to drive down the need for labour hours (decreasing the denominator) with significant intensity, thereby increasing the capital-to-labour ratio and precipitating increases in labour underutilisation. We posit the existence of an AI-capital-to-labour ratio threshold beyond which a self-reinforcing cycle of recessionary pressures could be triggered by reductions in disposable income and consumption brought on by too great an increase in labour underutilisation, and requiring sustained government intervention to maintain stability \cite{RN15}. Employing a system dynamics model, the aim of this research was to answer three primary research questions: (i) What are the possible trajectories for per capita disposable income and consumption / demand in the economy over the longer term in response to varying degrees of shifts in the capital-to-labour ratio; (ii) Could a capital-to-labour ratio threshold exist in the context of plausible parameter estimates, beyond which the economy may enter a self-reinforcing cycle of recessionary pressures, and if so, what are the conditions under which this might occur; and (iii) what level of new job creation would be required to prevent an overall reduction in consumption across the economy in the advent of the speculated substitution of 25\% of current work by generative AI \cite{RN19,RN20}. 

\label{Method }
\section*{Method }

\textbf{\textit{Model structure and assumptions}}

\begin{figure*}[!htb]
\centering
\includegraphics[width=1\textwidth]{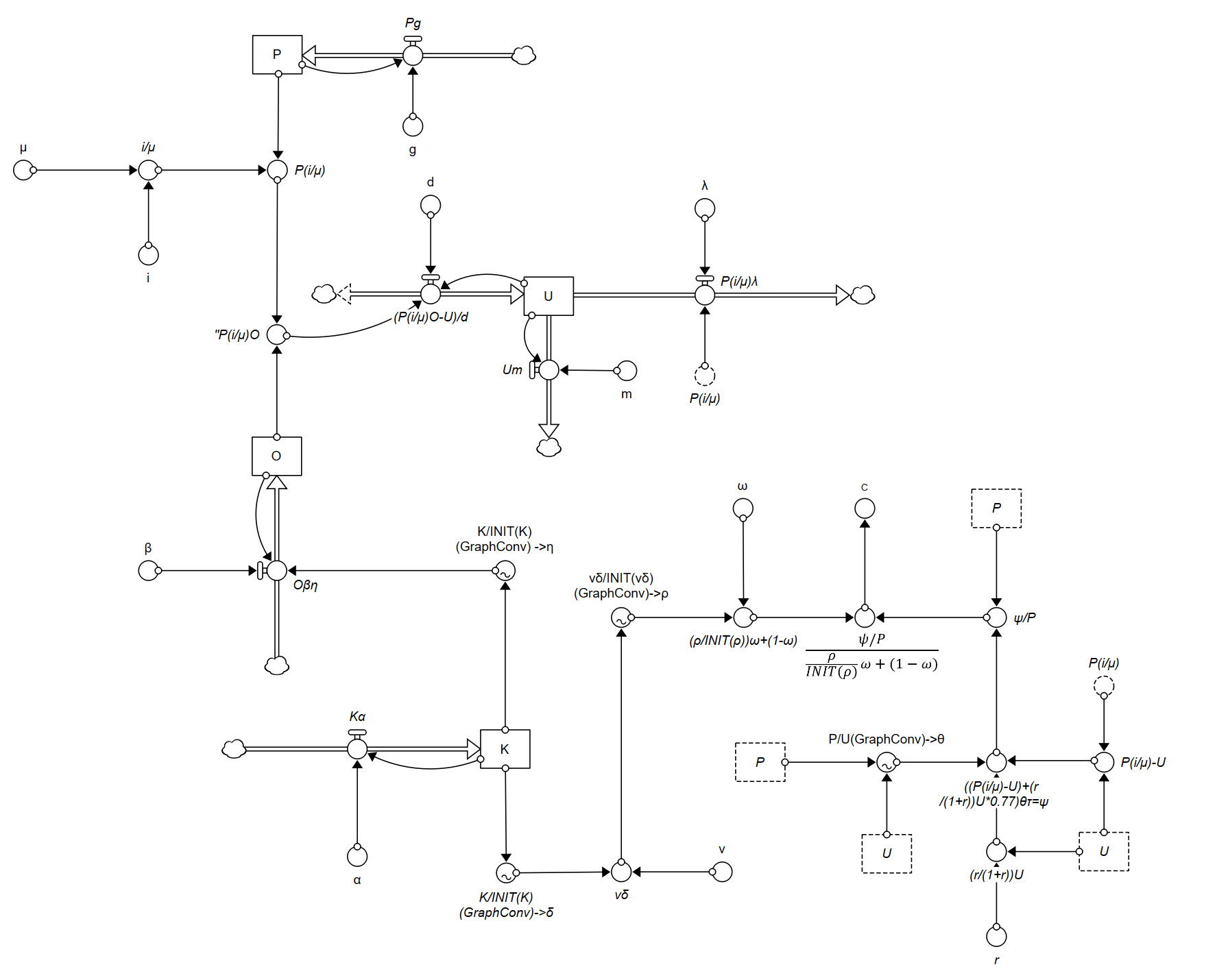} 
\caption{A system dynamics model of the interaction between the capital-to-labour ratio, labour underutilization, disposable income, and relative prices, and their impact on consumption. Symbols, their values, and data sources are defined in Appendix 1.}
\label{fig:image2}
\end{figure*}

Figure \ref{fig:image2} presents the system dynamics model used for the analyses. The model consists of four stocks (or state variables); $P, U, O$ and $K$ corresponding to population, the number of people who are unemployed or underemployed in the population (underutilisation), the labour underutilisation onset rate, and the capital-labour ratio (K-L ratio) respectively. People are added to the total population, $P$, at a constant fraction $g$ per year representing a net population increase (births plus immigration minus deaths and emigration) consistent with current population projections. Those becoming underutilised flow into the stock, $U$, at a rate (per year) equal to $\frac{P\frac{i}{\mu}O - U}{d}$, where $i$ is the number of people in the labour force in 2023.5, $\mu$ is the population in 2023.5, with $\frac{i}{\mu}$ giving the proportion of the population in the labour force, and $d$ is the delay in changes to the underutilisation onset rate as a result of changes to the K-L ratio trend. For simplicity, we assume that the proportion of the population in the labour force (participating in or seeking work) remains constant (i.e. this model does not account for major demographic shifts that would alter the proportion in the labour force within the model time horizon (mid-2050). People flow out of the underutilisation stock at a rate equal to $P\frac{i}{\mu}\lambda$, where $\lambda$ is the new job creation rate per capita (which assumes that these new jobs provide sufficient hours to move individuals out of a state of being unemployed or underemployed). All-cause mortality reduces the number of people unemployed and underemployed (underutilised persons) at a rate equal to $Um$, where $m$ is the per capita mortality rate for those of working age.

Labour underutilisation onset increases over time at a rate equal to $O\beta \eta$, where $\beta$ is the base rate of change in the labour underutilisation onset rate, and $\eta$ is the per unit change in the underutilisation onset rate as a result of changes to the K-L ratio (dependent on a graphical converter, see Appendix \ref{tab:input}). The K-L ratio increases at a rate per year equal to $K\alpha$, where $\alpha$ is the percent increase in the K-L ratio based on the historic average. Scenarios exploring AI-related increases to the K-L ratio (capital deepening) occur through changes to the value of $\alpha$. Multifactor productivity (MFP) changes at a rate equal to $\nu\delta$, where $\nu$ is the base MFP rate and $\delta$ is the per unit change in the MFP rate as a result of changes to the K-L ratio (dependent on a graphical converter, Appendix 1). Prices of goods and services, $\rho$, decrease as MFP increases (dependent on a graphical converter, Appendix 1). The impact of labour underutilisation on aggregate disposable income, ${\psi}$, is 
\[
\psi = (P\frac{i}{\mu}-U+\frac{r}{1+r}U \times 0.77) \theta \tau
\]

\afterpage{
    \clearpage
    \begin{table*}[ht]
    \centering
   \captionsetup{size=small} 
\caption{The simulated impacts of a range of possible intensities of generative AI-related capital deepening on labour underutilisation and proxies for demand.}
    \label{tab:results_summary}
\small
    \begin{tabular}{@{}lccccr@{}} 
    \toprule
Scenario & \makecell{Mean reduction\\against baseline}  & \makecell{Mean \%\\reduction} & \makecell{Median \%\\reduction} & \makecell{Lower 95\%\\interval} & \makecell{Upper 95\%\\interval} \\
\midrule
    \multicolumn{6}{c}{ Disposable income per capita USD} \\
    \midrule
    a.  K-L ratio 4\% increase per annum &  5,035.1 & 12.74 & 12.73 & 7.29 & 18.06 \\
    b.  K-L ratio 7\% increase per annum & 10,244.8 & 25.96 & 25.82 & 20.61 & 31.76\\
    c.  K-L ratio 10\% increase per annum & 12,630.6 & 25.96 & 32.06 & 26.06 & 37.84 \\
    \midrule
    \multicolumn{6}{c}{Consumption index} \\
    \midrule
    a.  K-L ratio 4\% increase per annum & 0.0674 & 7.34 & 7.59 & -0.23 & 14.79\\
    b.  K-L ratio 7\% increase per annum & 0.1939 & 21.21 & 21.03 & 13.56 & 28.33\\
    c.  K-L ratio 10\% increase per annum & 0.2527 & 27.66 & 27.92 & 20.14 & 34.80 \\
    \midrule
    \multicolumn{6}{c}{Underutilised persons} \\
    \midrule
    a.  K-L ratio 4\% increase per annum & 1,036,650 & 37.63 & 36.76 & 21.03 & 57.97 \\
    b.  K-L ratio 7\% increase per annum & 2,758,013 & 99.76 & 98.38 & 70.61 & 137.52 \\
    c.  K-L ratio 10\% increase per annum & 3,808,004 & 137.69 & 136.99 & 98.13 & 179.70 \\
    \bottomrule
    \label{tab:t1}
    \end{tabular}
    \end{table*}
}

where $r$ is the ratio of underemployed to unemployed, $\theta$ is the per unit change in disposable income as the proportion of the population underutilised changes (i.e. as the proportion of the population that is unemployed and underemployed increases, it is assumed to exert a downward pressure on wages and hence disposable income - dependent on a graphical converter, Appendix 1), and $\tau$ is the scaling factor derived by calibration that ensures the model’s initial conditions broadly reflect real world disposable income (namely, Real Net National Disposable Income, a measure of real income available across the economy to spend or save). The scaling factor takes account of the disposable income that is unlikely to be affected by downward pressure on wages (e.g. social security benefits, investment income, etc.).  The per capita disposable income is given by $\frac{\psi}{P}$.

The consumption index, $C$ is the per capita ability to purchase goods and services in the economy that accounts for productivity-related price decreases due to generative AI, and is equal to 
\[
\frac{\psi/P}{\frac{\rho}{INIT(\rho)} \omega+(1-\omega)}
\] 
where $\omega$ is the proportion of disposable income affected by productivity-related price decreases (it is assumed, for example, that increased productivity in the economy will not decrease the cost of housing, education, insurance, etc), and ‘INIT’ refers to the initial value of a parameter at the start of a simulation (so $\rho/INIT(\rho)$ is the ratio of current prices to initial prices and assesses how much prices change over time as a proportion of the original value). Therefore, the consumption index reflects the competing forces of productivity-related price reductions (which increase real disposable income) and reductions in disposable income due to increased underutilisation and downward pressure on wages.

\label{Model analysis}
\section*{Model analysis}
Australian data was used to derive plausible parameter estimates for the model as a demonstration of the feasibility of the hypothesis of recessionary pressures associated with scaled uptake of generative AI across the economy. The potential impacts of generative-AI were modelled through changes to $\alpha$ (the percent increase in the K-L ratio, also known as capital deepening). The baseline scenario reflects the average increase in $\alpha$ of 1.8\% per annum derived from data from the Australian Bureau of Statistics (ABS) over the period 1995 to 2023. To address research question (i), a range of possible intensities of capital deepening through uptake of generative AI across the economy (i.e. increases to the AI-capital-to-labour-ratio) were modelled through increases in the value of $\alpha$ to 4\%, 7\% and 10\% per annum (scenarios a, b, and c respectively). These K-L ratio scenarios are conservative as the 10\% increase per annum is projected to increase underutilization by 23.8\% by mid-2050 (assuming no government action), which is below the estimated 25\% of work/jobs that will be affected across the economy as reported by OpenAI researchers, Goldman Sachs, and others \cite{RN19, RN20, RN21}.  Impacts on several key outcomes were explored at the end of the simulation time horizon (mid-2050); per capita disposable income (expressed in US dollars) and the consumption index (proxies for demand in the economy) and labour underutilisation. Sensitivity analyses were performed to assess the impact of uncertainty in estimates of the new job creation rate, the proportion of disposable income subject to productivity-related price decreases, the base labour underutilisation onset rate increase per year, and the ratio of underemployed to unemployed on the simulation results.  We used Latin hypercube sampling to draw 200 sets of values for the selected parameters from a uniform joint distribution (the sample space for each parameter is provided in Appendix 1). Given that new job creation above the baseline rate (whether occurring through new industries emerging from the generative AI-related technical disruption or through government stimulus) is unlikely to occur instantaneously, a 2-year delay was implemented with an s-shaped onset over time.  Differences in projected outcomes between the baseline and scenario runs were calculated for each set of parameter values and summarised using simple descriptive statistics.  Model construction and sensitivity analyses were performed using Stella Architect ver. 3.4 (\url{www.iseesystems.com}). Microsoft Excel, R, and Python were used for post dynamic modelling summary statistics and graphical presentation of results.

To address research question (ii), a sensitivity analysis was performed using all possible combinations of the annual K-L ratio percent increase (ranging from 2\% to 10\%; increments of 0.28) and new job creation rate (ranging from a 0.5- to 12-fold increase in the historic average of the new job creation rate; increments of 0.4), drawing 900 sets of values from a joint distribution incrementally evenly distributed. A heatmap was generated to visualise the conditions under which the simulated consumption index would fall below the baseline trajectory. To address research question (iii), the K-L ratio was set to an 11\% increase per annum which approximates the speculated substitution of 25\% of current work/jobs by generative AI. A range of values of an increase in the new job creation rate were explored (i.e. between a 1- to 12-fold increase) to understand the extent to which new jobs would need to be created to prevent a decline in the consumption index compared to the baseline over the next 20 years (2025-2045) in the absence of government intervention.

\begin{figure*}[!htb]
\centering
\includegraphics[width=1\textwidth]{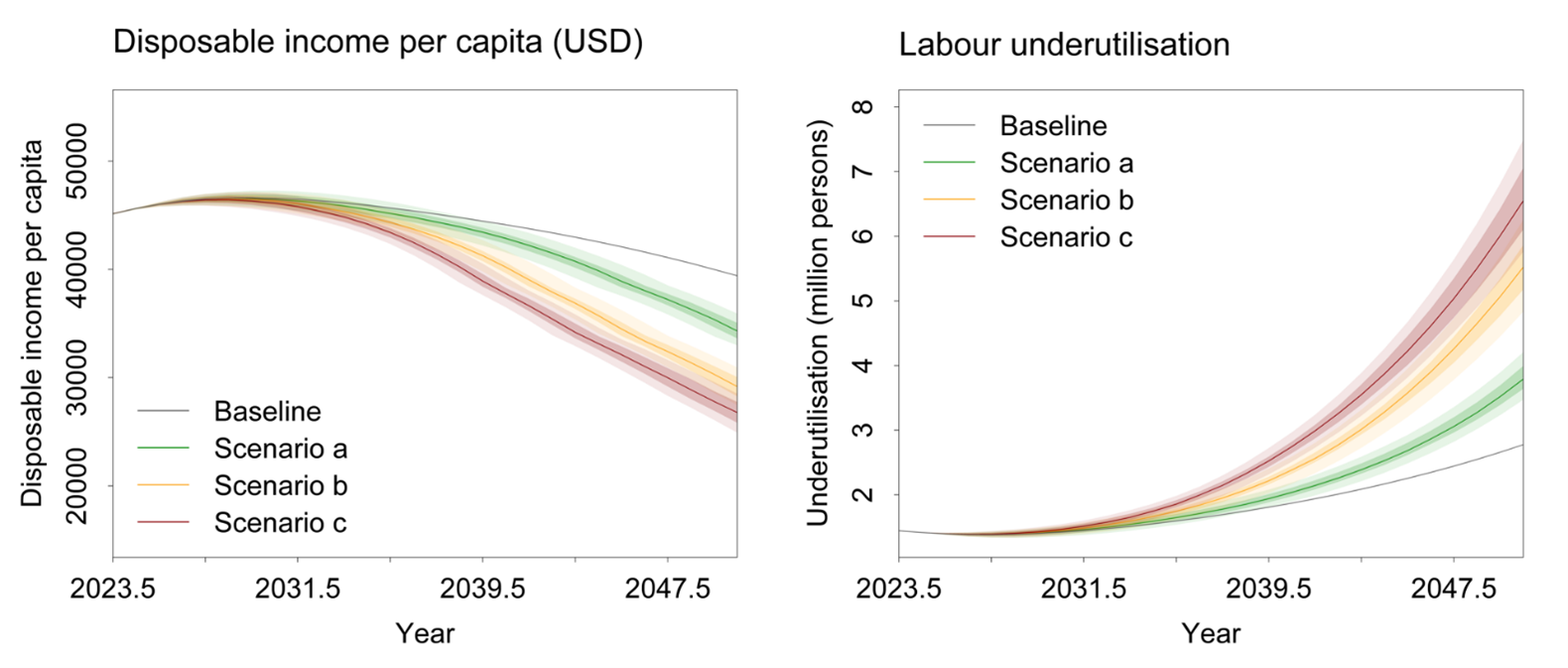}
\caption{Simulation results. Disposable income per capita in US dollars and underutilised persons for alternative scenarios between 2023.5 and 2050.5, compared with the baseline. The baseline and scenario’s a, b, and c reflect an average increase in the AI-capital-to-labour ratio of 1.8\%, 4\%, 7\%, and 10\% per annum respectively. The solid line indicates the simulated median values over time, darker shading represents 50\% uncertainty interval, and lighter shading indicates the 95\% uncertainty interval derived from the sensitivity analysis. Declining disposable income per capita in the baseline scenario reflects the increasing trend in the underutilisation rate.}
\label{fig:image3}
\end{figure*}

\label{Results}
\section*{Results}

\begin{figure}[!htb]
\centering
\includegraphics[width=0.48\textwidth]{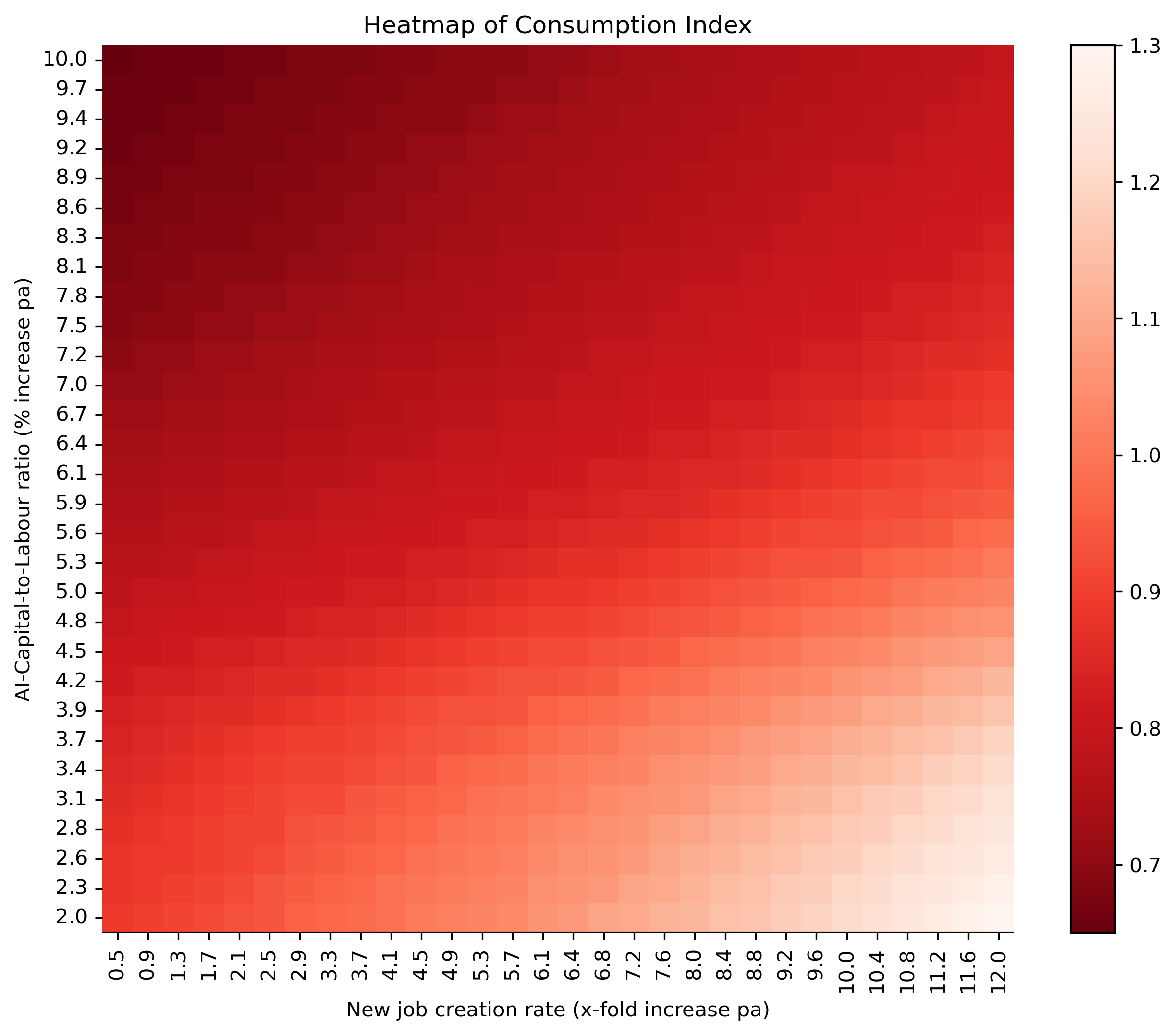}
\caption{Simulation results.  The heatmap shows the impacts of the AI-capital-to-labour ratio and new job creation rate on the consumption index over the period 2023.5 to 2050.5. The pale shading corresponds to increases in the consumption index against the baseline, the lighter red shading indicates smaller percentage reductions against the baseline, and dark red shading to larger percentage reductions.}
\label{fig:image4}
\end{figure}

\begin{figure}[!htb]
\centering
\includegraphics[width=0.48\textwidth]{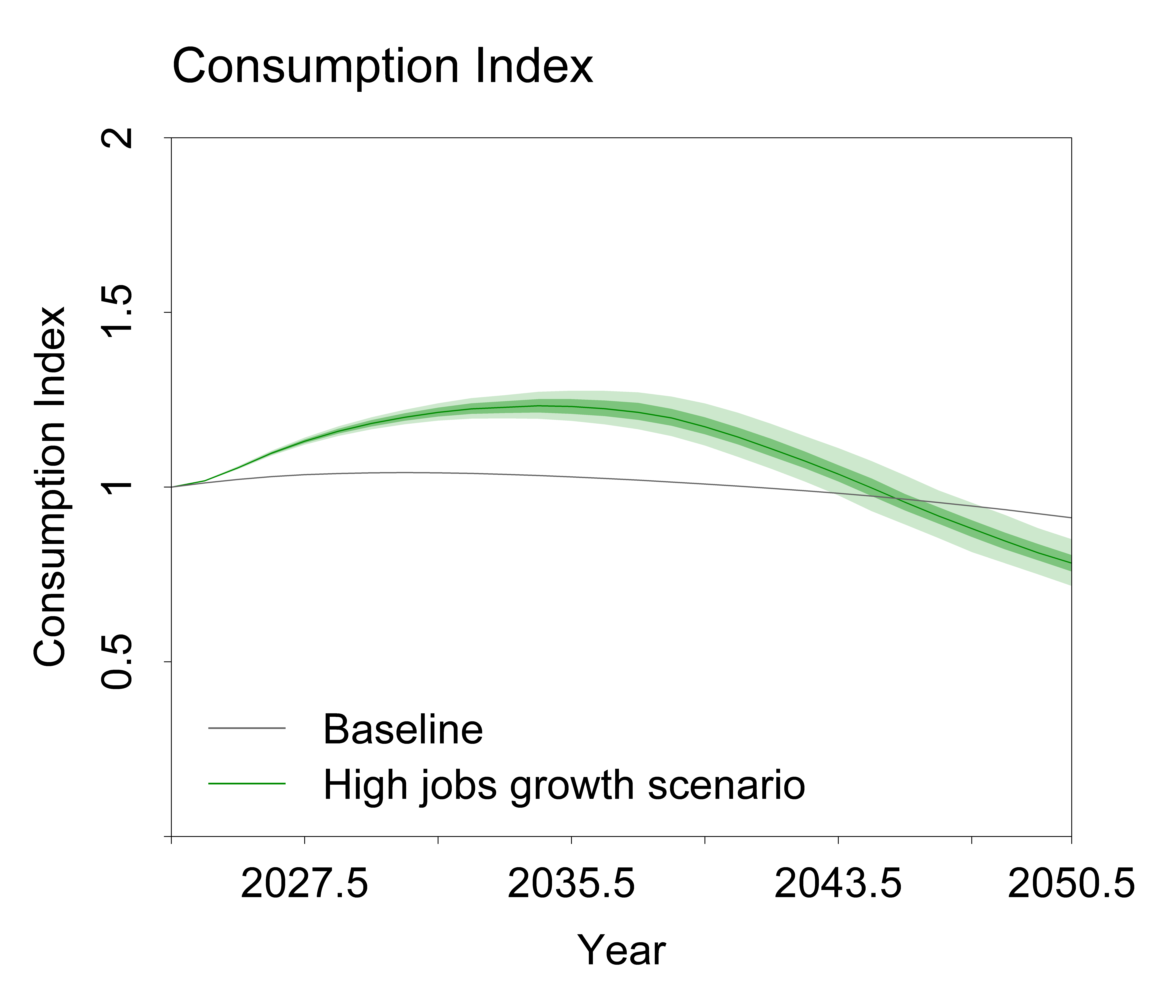}
\caption{Simulation results. The figure shows the non-linear change in consumption index between 2023.5 and 2050.5 under a scenario of a moderate rate of AI-capital deepening (K-L ratio increase per annum of 7\%) and high rate of jobs growth (a 6-fold increase in the baseline new job creation rate per annum) compared to the baseline. The solid line indicates the simulated consumption index over time, darker shading represents 50\% uncertainty interval, and lighter shading indicates the 95\% uncertainty interval derived from the sensitivity analysis.}
\label{fig:image5}
\end{figure}

Regarding research question (i), Table 1 shows simulation results for alternative scenarios of generative AI-related capital deepening on per capita disposable income, the consumption index, and underutilised persons against the baseline over the period 2023.5 to 2050.5.  Scenario b which signifies a moderate increase in the historic average annual increase in the K-L ratio due to generative AI, is projected to result in a mean 99.76\% (95\% uncertainty interval, 70.61\% – 137.52\%) increase in underutilisation, a 21.2\% (95\% interval, 13.6\% - 28.3\%) decrease in the consumption index, and a 26\% (95\% interval, 20.6\% - 31.8\%) decrease in the disposable income per capita (note that all intervals reported are derived from the distributions of model outputs calculated in the sensitivity analyses; they provide an indication of the effect of uncertainty in selected parameter estimates but should not be interpreted as confidence intervals).  The simulated trajectories provided in Figure \ref{fig:image3} highlight that the intensity of generative AI-related capital deepening is likely to influence the timing and extent of the increase in labour underutilisation and declines in disposable income per capita over the longer term.
Regarding research question (ii), Figure \ref{fig:image4} demonstrates the values of the consumption index (the ability to purchase goods and services in the economy) as a function of the AI-capital-to-labour ratio and rate of new job creation (either through innovation in the market or government stimulus). The figure demonstrates a threshold beyond which increases in the AI-capital-to-labour ratio would result in values of the consumption index that fall below the baseline over the longer term, and that the threshold depends on the rate of new job creation. At low levels of increase in the AI-capital-to-labour ratio, increases in the new job creation rate is sufficient to prevent downturns in the consumption index. However, as demonstrated in Figure \ref{fig:image5}, at moderate to high levels of the AI-capital-to-labour ratio, while initially the consumption index increases against the baseline, even high rates of growth in the new job creation rate are insufficient to offset labour underutilisation and associated declines in disposable income per capita. This suggests that the AI-capital-to-labour ratio represents a potential economic marker for threats to national wellbeing that could be closely monitored. Regarding research question (iii), the model estimates (in this example) that at least a 10.8-fold increase in the current rate of new job creation would be needed to prevent a decline in the consumption index compared to the baseline over the next 20 years (2025-2045) in the absence of government intervention (recalling that the model assumes that the new jobs will provide sufficient hours to move individuals out of a state of being unemployed or underemployed). However, a far higher growth rate in new jobs would be needed to achieve and sustain growth in demand above the baseline.

\label{Discussion}
\section*{Discussion}
The aim of this study was to assess the merit of the hypothesis that recessionary pressures might be associated with expanded implementation of generative AI across the economy, employing plausible parameter estimates.  The findings suggest that without government intervention, generative AI has the potential to significantly disrupt labour market dynamics and demand in an economy emphasizing the need for proactive policy measures to mitigate adverse impacts. Projections highlight the risk of substantial increases in labour underutilization even from moderate levels of AI-capital deepening, leading to declines in per capita disposable income and consumption despite increased consumer purchasing power related to productivity growth and lowering of prices. Furthermore, our study suggests the existence of an AI-capital-to-labour ratio threshold beyond which the economy may enter a self-reinforcing cycle of recessionary pressures. Such a scenario could pose formidable challenges to economic stability and necessitate sustained government intervention to prevent social and economic instability. The AI-capital-to-labour ratio therefore represents a potential economic marker for threats to economic and social wellbeing in the Age of Intelligence. The findings underscore the urgency of developing strategies to manage the transition towards an AI-dominated economy while safeguarding Mental Wealth and social prosperity.

While these analyses were conducted within a specific context (Australia), and not designed to precisely determine where the threshold in the AI-capital-to-labour ratio lies, the results indicate that such exploration should be conducted across a variety of contexts. Whether a threshold exists and where it lies will depend on the features of each economy including industry composition, occupational structures, tax structures, the regulatory environment, government policies and incentives, and social acceptability and environmental considerations of the integration of generative AI in workplaces and everyday life. Conducting risk mapping based on AI-capital-to-labour ratio and other indicators across different economy archetypes may allow for the creation of tailored strategies to manage the transition to an AI-augmented economic landscape, while recognizing that these thresholds may be dynamic as the features of economies evolve. Advanced economies with a high degree of automation and digital infrastructure may face different challenges compared to emerging markets where labour-intensive industries still predominate and there is a shortage of skilled workforces to capitalize on the advantages of generative AI \cite{RN22}. Similarly, economies with robust social institutions and safety nets may have different risk profiles compared to those with less comprehensive support systems \cite{RN23}. Economic inter-dependencies and global trade dynamics can also shape the trajectory of AI-capital deepening. Economies that rely heavily on exports or are integrated into global supply chains may face pressure to adopt generative AI technologies to remain competitive, potentially accelerating the approach to the threshold. Further empirical work is clearly needed; however, by closely tracking changes in this ratio, both within sectors and across the economy, governments can better anticipate and respond to emerging challenges, thereby enhancing their capacity to ensure economic resilience and social cohesion.

\vspace{1em}
\textbf{\textit{Pre-distribution: The role of governments \& labour unions}}

While beyond the scope of this aggregate-level analysis, increases in labour underutilisation are likely to bring labour market reshaping, along with the risk of further deepening wage polarization and inequality \cite{RN15, RN24, RN25}. As generative AI technologies advance, enabling greater productivity with less human input in non-routine cognitive or professional jobs, the distribution of economic gains will become further skewed and the disproportionate accrual of economic benefits to the owners of capital is likely to aggravate existing wealth disparities across generations \cite{RN26,RN27,RN28,RN29}. Young people entering the workforce at this time will face significant challenges. The automation of entry-level professional positions—the bedrock for skill-building and early career development—threatens to undermine young workers’ ability to establish themselves economically. The absence of these foundational roles may impede the accumulation of wealth for this generation, potentially leading to a widening intergenerational wealth gap as they face a labour market with fewer quality jobs and stagnant wages \cite{RN28}. This shift not only affects their immediate employment prospects but also has repercussions for long-term financial stability and prospects for wealth creation. The erosion of stable employment and career progression, fundamental to young adults’ aspirations and well-being, increases the likelihood that young people will go on to experience high psychological distress and mental disorder, increasing the risk of suicidal behaviours and undermining social cohesion over the longer term \cite{RN30, RN31}. This further underscores the need for proactive measures to support all workers, but particularly young workers, and ensure that the path to prosperity remains accessible and inclusive in the Age of Intelligence. 

In light of these challenges, redistribution measures such as taxation and progressive social policy have been proposed as a means to addressing widening inequalities associated with generative AI-associated job scarcity, displacement, and downward pressure on wages \cite{RN15, RN32}. However, while redistribution efforts can play a role in alleviating adverse impacts of generative AI, such measures are reactive and retain systemic issues already in place \cite{RN33}. Additionally, history has demonstrated that even in the presence of abundance, perceived scarcity persists, and the struggle for redistribution of wealth can incite conflict and division \cite{RN16, RN28}. A more forward-looking strategy lies in the concept of pre-distribution, which seeks to address the root causes of inequality by adjusting the mechanisms of wealth generation before inequalities become entrenched \cite{RN34}. Pre-distribution policies focus on ensuring that the economic structures and market forces themselves generate equitable outcomes, rather than relying solely on redistribution. 

Governments will need to play a crucial role in shaping the landscape for pre-distribution in the Age of Intelligence. Strategies could include investing in the foundational economy and underfunded sectors with strong job creation potential, such as health, social care, science, arts and culture, and the green economy. Exploring public policies such as reduced workdays with maintained salaries to ensure that the benefits of AI-induced productivity are translated into widespread well-being \cite{RN32}. Additionally, governments can encourage private sector compliance with ethical AI adoption standards, through incentivizing businesses to comply with initiatives like an AI Fairwork Pledge \cite{RN32}. Innovative mechanisms such as an AI Displacement Insurance Fund \cite{RN32} are another avenue for government action, where public-private collaboration would see companies that automate jobs contribute to a pool of resources dedicated to easing labour market transitions and supporting workers affected by AI-driven changes \cite{RN32}. Through multifaceted strategies, governments, in partnership with civic organisations can lay the foundation for a more inclusive economy, where the transformative potential of AI is harnessed for the good of all, rather than exacerbating existing divides. By prioritizing pre-distribution, policymakers can address disparities at their source, fostering a society that is resilient, cohesive, and prepared for the future.

Similarly, labour unions have a critical role to play in this rapidly evolving economic landscape shaped by generative AI, in promoting fair labour practices, negotiating job security, and lobbying for pre-distribution policies that ensure the gains from AI-driven productivity are equitably shared. Unions are in a unique position to influence policy discussions around generative AI, safeguarding the interests of a new generation of workers and preventing their marginalisation. Tripartite planning involving unions, government, and employers can facilitate solutions that respect the dignity of labour, support workers’ mental health, and preserve the social fabric of nations. To fulfill this role, unions could work with government and employer representatives to develop an international labour standard on ‘Decent work in the Age of Intelligence,’ ensuring that AI-related jobs maintain high-quality standards and are incorporated into national legislation \cite{RN32}. Establishing such labour standards will assist in negotiating for AI-related work that augments, rather than substitutes, human intelligence (brain capital), creativity, and innovation, pushing for clauses in employment contracts or industry-wide agreements that set ethical standards for AI use, include protections against unjustified job displacement and ensure fair compensation for AI-generated work. Additionally, unions could advocate for the establishment of regulatory frameworks for data-knowledge ownership (e.g., data cooperatives or trusts \cite{RN35, RN36}) that ensure individuals are fairly compensated and retain control over their data and intellectual contributions to the development and refinement of generative AI. Through such actions, unions can position workers at the centre of technological governance to positively shape labour market evolution and avoid the destructive trajectory of technological determinism. 

The recent triumph of the Writers Guild of America (WGA) over the studios’ attempted use of generative AI to automate scriptwriting serves as a beacon for labour unions navigating the complexities of generative AI in the workplace. This victory, however, does not guarantee future success as generative AI becomes more subtly woven into the fabric of work. Unions are now operating in a terrain vastly different from the one in which they originated, necessitating a fundamental rethink of their strategies. As unions like the WGA demonstrate, there is significant power in collective resistance against the exploitative use of generative AI. However, in this new labour market landscape, unions must adapt and forge a balanced approach that fosters innovation and business opportunities while upholding hard-won workers’ rights and incomes. To achieve this equilibrium, unions must actively engage in AI governance, ensuring representation in critical discussions with employers, policymakers, and technologists, on the adoption of AI in the workplace, and push for transparency and ethical standards. A comprehensive mapping of professions and sectors, including supply chain and distribution networks, could equip labour unions to engage in strategic, pre-emptive measures to safeguard vulnerable jobs. Additionally, educational partnerships will be crucial, ensuring young workers have access to training programs that equip them with the skills needed for AI-augmented jobs. Transition to a generative-AI powered economy presents a unique opportunity to reevaluate and renew the social contract between employers, employees, government, and broader society. This period of transition offers a significant opportunity for labour unions to demonstrate their enduring value as an essential pillar for work justice and human rights. This subtle yet pivotal moment could mark a renaissance for the union movement, affirming its role in a fair and forward-looking labour market.

\label{Study_limit}
\section*{Study limitations}
It is crucial to recognize the inherent uncertainties and constraints associated with any modelling effort. The complexities of the economy and the interactions among myriad factors pose a challenge to comprehensively encapsulate all elements within a single model. Nonetheless, striving for an exhaustive model is neither necessary nor particularly beneficial. A simplified, abstract, and transparent systems model can provide profound insights into the dominant forces shaping complex system behaviours. A limitation of our current approach is its focus on macroeconomic indicators and does not have the capacity to explore the differential impacts across various demographic groups, sectors, economies, or income levels. For example, while the model projects a decline in disposable income, it does not delineate how this decline may disproportionately impact different income strata. Additionally, it does not make a distinction between high-skilled, well-compensated jobs and lower-skilled less remunerative ones in simulating alternative rates of new job creation. This may obscure the full extent of the impact of generative AI on labour underutilisation, disposable income, and overall economic demand. However, it is important to reiterate the study’s intention which was to demonstrate the plausibility of a threshold in the AI-capital-to-labour ratio existing, thereby validating the hypothesis.
Future research should aim to refine and broaden the modelling framework to account for a wider array of factors and nuances. This includes examining the varied impacts of generative AI adoption across different industries and in diverse global economic settings, as well as the influence of policy measures in determining these outcomes. It is also important to consider potential feedback loops that intertwine economic and social variables. Empirical studies that support risk mapping, validate model predictions, and scrutinize the effects of real-world policy applications will be instrumental in fostering evidence-based policy making, especially as we navigate the implications of this major technological advance.

\setcounter{table}{0}
\captionsetup[table]{name=Appendix} 

\begin{table*}[ht]
\centering
 \captionsetup{size=small}
\caption{Parameter values used in the model analysis and data sources.}
\label{tab:input}
\scriptsize
\begin{tabular}{@{}lllcr@{}} 
\toprule
\setlength{\heavyrulewidth}{1pt}
Parameter & Symbol & Value & Data source \\ 
\toprule
\makecell[l]{Underutilised persons initial} & $U_{0}$ & 1,445,000 & \makecell[l]{Unemployed plus underemployed persons in Australia as of 30th June 2023 – seasonally \\adjusted ABS, 6202.0 Labour Force, Australia, Table 22.  https://www.abs.gov.au/statistics/\\labour/employment-and-unemployment/labour-force-australia/latest-release) } \\

\hline

\makecell[l]{Ratio of underemployed\\ to unemployed} & \textit{r} & 1.6 & \makecell[l]{Derived from ABS, 6202.0 Labour Force, Australia, Table 22.\\ https://www.abs.gov.au/statistics/labour/employment-and-unemployment/\\labour-force-australia/latest-release. Average value over the 10-year period,\\ Dec 2013 – Nov 2023. Varied by +/-10\% in sensitivity analyses.}\\

\hline

\makecell[l]{Ratio of disposable income\\ of an underemployed to\\ fully employed person} & - & 77\% & \makecell[l]{Derived from Table 25 of: Campbell I., Parkinson, S. and Wood, G. (2014) \\Underemployment and housing insecurity: an empirical analysis of HILDA data, AHURI \\Final Report No.230. Melbourne: Australian Housing and UrbanResearch Institute.} \\

\hline

\makecell[l]{Labour underutilisation onset\\ rate initial} & $O_{0}$ & 9.9\% &\makecell[l]{ Derived from ABS, 6202.0 Labour Force, Australia, Table 22 \\https://www.abs.gov.au/statistics/labour/employment-and-unemployent/\\labour-force-australia/latest-release} \\

\hline

\makecell[l]{Base labour underutilisation\\ onset rate increase per year} & $\beta$ & 0.0015 & \makecell[l]{ Reasoned estimate based on plausible forward projections of labour\\ underutilisation to 2050.5. Varied by +/-10\% in sensitivity analyses.} \\

\hline

\makecell[l]{Labour force initial} & \textit{i} & 14,585,316 & \makecell[l]{Labour force total persons as of 30th June 2023. ABS, 6202.0 Labour Force, \\Australia, Table 22: https://www.abs.gov.au/statistics/labour/employment-and-\\unemployment/labour-force-australia/latest-release. } \\

\hline

\makecell[l]{Capital-to-labour ratio\\ (K-L ratio) initial} & $K_{0}$ & 94.6 & \makecell[l]{ABS, Australian National Accounts, 2022-23, Cat. no. 5204.0, Table 13: https://www.abs.gov.au\\/statistics/economy/national-accounts/australian-system-national-accounts/\\latest-release. Quality adjusted hours worked and indexed (2021=100)} \\

\hline

\makecell[l]{K-L ratio percent increase\\ per annum} & $\alpha$ & 1.8\% & \makecell[l]{ABS, Australian National Accounts, 2022-23, Cat. no. 5204.0, Table 13:\\ https://www.abs.gov.au/statistics/economy/national-accounts/australian-system\\-national-accounts/latest-release. Average annual increase in the K-L ratio \\over the period 1995 to 2023 derived from regression model.}\\

\hline

\makecell[l]{Initial population} & $P_{0}$ & 26,638,544 & \makecell[l]{The population of Australia as of 30th June 2023 \\(https://www.abs.gov.au/statistics/people/population) }\\

\hline

\makecell[l]{Population increase percent\\ per annum} & $g$ & 1.1 & \makecell[l]{Population projections, Australia:https://www.abs.gov.au/statistics/people\\/population/population-projections-australia/2022-base-2071. This approximates\\ medium series by 2050.5.}\\

\hline

\makecell[l]{Mortality rate} & \textit{m} & 0.0015 & \makecell[l]{Derived from Australian Bureau of Statistics. Table 1.9, Life Tables, \\Australia, 2020-2022 https://www.abs.gov.au/statistics/people/population/life-\\expectancy/latest-release and population data for each age.}\\

\hline

\makecell[l]{New job creation rate per\\ capita} & $\lambda$ & 0.0021 & \makecell[l]{To align the model`s output with the June 2023 Real Net National Disposable Income\\ (RNNDI) per capita in Australia, a scaling factor of 86,985 was applied. This calibration ensures\\ the model`s initial conditions broadly reflect a real world economic indicator of disposable\\ income (data: Table 1: https://www.abs.gov.au/statistics/economy/national-accounts/australian\\-national-accounts-national-income-expenditure-and-product/latest-release\#data-downloads)}\\

\hline

\makecell[l]{Delay in change to \\underutilisation onset rate} & \textit{d} & 5 & \makecell[l]{5 years – an estimate. Delay in changes to the underutilisation onset rate \\as a result of changes to the K-L ratio.} \\

\hline

\makecell[l]{Base Multi-Factor \\Productivity (MFP) rate} & $v$ & 0.56298 & \makecell[l]{Derived from MFP data, OECD, Australia, 1985-2022:\\ https://data.oecd.org/lprdty/multifactor-productivity.htm}\\

\hline

\makecell[l]{Scaling factor} & $\tau$ & 86,985 & \makecell[l]{Derived from MFP data, OECD, Australia, 1985-2022:\\ https://data.oecd.org/lprdty/multifactor-productivity.htm}\\

\hline

\makecell[l]{Proportion of disposable income\\ affected by productivity-related\\ price decreases} & $\omega$ & 50\%& \makecell[l]{Reasoned estimate. Varied between 30\% and 70\% in sensitivity analyses.}\\

\hline

\makecell[l]{The per unit change \\in average disposable income\\ as the proportion of \\the population underutilised\\ changes} & $\theta$ & - & 
\makecell[l]{Graphical converter: Points (underutilisation, disposable income): \\(0.000, 1.359), (0.500, 1.152), (1.000, 1.000), (1.500, 0.876), (2.000, 0.796),\\ (2.500, 0.748), (3.000, 0.705), (3.500, 0.676), (4.000, 0.648), (4.500, 0.631), (5.000, 0.612).\\ Value of 1 is reference point (i.e. current values of underutilisation  and average \\ disposable income). Highlighted example says that if underutilisation was double \\ the current rate, then average disposable income (wages) would be 79.6\% the current level \\ (or 20.4\% less than the current level). Approximates Phillips Curve provided by the Reserve Bank \\of Australia (Figure 3 in https://www.rba.gov.au/publications/rdp/2021/pdf/rdp2021-09.pdf) }\\

\hline

\makecell[l]{The per unit change in the\\ underutilisation onset rate\\ as a result of changes\\ to the K-L ratio due to\\ generative AI} & $\eta$ & - & \makecell[l]{Graphical converter: Points (K-L ratio, labour underutilisation onset rate): (0.000, 0.000),\\ (0.500, 0.291), (1.000, 1.000), (1.500, 1.990), (2.000, 3.083), (2.500, 4.029), (3.000, 4.636),\\ (3.500, 4.976), (4.000, 5.000), (4.500, 5.000), (5.000, 5.000).}\\

\hline

\makecell[l]{The per unit change in\\ the MFP rate as a result of\\ changes to the K-L ratio\\ due to generative AI} & $\gamma$ & - & \makecell[l]{Graphical converter: Points (K-L ratio, MFP): (0.000, 0.000), (0.500, 0.415),\\ (1.000, 1.000), (1.500, 1.524), (2.000, 2.214), (2.500, 3.135), (3.000, 4.093),\\ (3.500, 4.667), (4.000, 4.929), (4.500, 5.000), (5.000, 5.000)}\\

\hline

\makecell[l]{The per unit change in \\prices as MFP increases} & $\rho$ & - & \makecell[l]{Graphical converter (MFP, prices): Points (MFP, Prices): (0.000, 1.3890),\\ (0.500, 1.1500), (1.000, 1.0000), (1.500, 0.9223), (2.000, 0.8870), (2.500, 0.8560),\\ (3.000, 0.8332), (3.500, 0.8083), (4.000, 0.8000), (4.500, 0.8000), (5.000, 0.8000) }\\
\bottomrule
\end{tabular}
\end{table*}

\label{Conclusion}
\section*{Conclusion}
This study reveals the paradox of generative AI’s impact on prosperity: while it holds the promise of increased productivity and wealth creation, it also poses a substantial risk to economic stability and societal well-being. Our results suggest the existence of a threshold in the AI-capital-to-labour ratio beyond which the pace of new job creation would likely be insufficient to offset the adverse effects of labour displacement. This threshold serves as a novel economic marker, a warning signal for policymakers to monitor closely as AI continues to advance. This study serves as a platform for a broader dialogue on the role of governments, industry, and unions in shaping a future where technological advancements are harnessed for the collective good. It is a call to action for strategic, informed, and coordinated efforts to manage a just transition toward an AI-augmented economy, ensuring that prosperity is not only preserved but also equitably shared, and that the Mental Wealth of nations is protected.

\subsection*{Funding and acknowledgements:}
This work was undertaken under the Mental Wealth Initiative supported by seed funding and philanthropic gifts provided to the Brain and Mind Centre, University of Sydney.

\subsection*{Author contribution:}
Manuscript concept and drafting: JO \& AP; Critical revision of manuscript and contribution of important intellectual content: all authors.

\appendix

\bibliographystyle{unsrt} 
\bibliography{AI_model.bib}







\end{document}